\documentclass{article}

\usepackage{arxiv}

\usepackage[utf8]{inputenc} % allow utf-8 input
\usepackage[T1]{fontenc}    % use 8-bit T1 fonts
\usepackage{hyperref}       % hyperlinks
\usepackage{url}            % simple URL typesetting
\usepackage{booktabs}       % professional-quality tables
\usepackage{amsfonts}       % blackboard math symbols
\usepackage{nicefrac}       % compact symbols for 1/2, etc.
\usepackage{microtype}      % microtypography
\usepackage{lipsum}		% Can be removed after putting your text content
\usepackage{graphicx}
\usepackage{doi}
\usepackage{amsmath}
\usepackage{authblk}
\usepackage[title]{appendix}
\usepackage[sort&compress,numbers]{natbib}
\title{Spray flamelet structures in a tubular counterflow configuration}

\author[1]{Francisco Rivadeneira}
\author[1]{Felipe Huenchuguala}
\author[2]{Arne Scholtissek}
\author[2]{Christian Hasse}
\author[3]{Eva Gutheil}
\author[ ~,1]{Hernan Olguin\thanks{Corresponding author\\ Email address: hernan.olguin@usm.cl}}

\affil[1]{Department of Mechanical Engineering, Universidad Técnica Federico Santa María,\linebreak Avenida España 1680, Valparaíso, Chile\linebreak}
\affil[2]{Institute for Simulation of reactive Thermo-Fluid Systems, TU Darmstadt,\linebreak Otto-Berndt-Stra{\ss}e 2, 64287 Darmstadt, Germany\linebreak}
\affil[3]{Interdisciplinary Center for Scientific Computing, Heidelberg University,\linebreak Im Neuenheimer Feld 205, 69120 Heidelberg, Germany}
\date{}

\begin{document}
\maketitle

\begin{abstract}
In this work, spray flamelet structures subject to curvature are systematically studied, emphasizing the ways in which this quantity modifies the budgets of the corresponding flamelet equations and their stretch-induced extinction limit. More specifically, a theoretical extension of the tubular counterflow configuration is first proposed, which allows the injection of droplets from the inner cylinder. After appropriate mathematical descriptions for this new configuration in physical and composition space are introduced, several ethanol/air tubular counterflow flames are studied. It is found that increasing curvature leads to major modifications of the resulting flamelet structures, which is attributable to its influence on the evaporation profiles. Further, it is found that increasing curvature considerably reduces the stretch-induced extinction limit, which can be directly related to a corresponding reduction of the maximum mixture fraction within the flamelet. Finally, it is concluded that extinction in tubular counterflow spray flames occurs through a mechanism significantly different from what has been previously observed for gas flamelets.
\end{abstract}

\vspace{15mm}

\section{Introduction}\label{introduction}

During the last decades, the tubular counterflow configuration has been introduced as a canonical framework for the systematic study of curvature effects in gas flames~\cite{Wehrmeyer01,Dixon-Lewis90,Dixon-Lewis91,Hu07,Hu2009,pitz2014,Xuan14,Scholtissek16,Wang06,Wang07}. This setup admits a one-dimensional mathematical description (which reduces to the classical equations for planar flames when curvature is set to zero) and it is established by issuing fuel from an inner porous cylinder towards an oxidizer that is injected from an outer porous tube, or vice versa~\cite{Wehrmeyer01,pitz2014,Scholtissek16}. Among the most important advantages of the configuration is that it allows the independent variation of stretch and curvature, which has facilitated carrying out parametric analyses illustrating the ways in which curvature modifies both flame structures~\cite{Hu07,Hu2009,pitz2014,Xuan14} and extinction limits~\cite{Hu07,Hu2009,pitz2014,Scholtissek16,Wang07}.

The positive features of the tubular counterflow configuration have also been relevant in the context of non-premixed gas flamelet theory~\cite{Xuan14,Scholtissek16}. It has been recently shown that, when differential diffusion is considered, the inclusion of curvature in the gas flamelet equations leads to additional convective contributions in composition space~\cite{Xuan14,Scholtissek16,Xu13,Scholtissek15,Scholtissek17}. The influence of these curvature terms has been evaluated in the context of different tubular non-premixed counterflow flames~\cite{Scholtissek16}, revealing that, while not very dominant in terms of relative budgets of the corresponding non-premixed flamelet equations, they could significantly modify flamelet structures. These results represent a major motivation for the study of curvature in more complex flames, such as spray flames. So far, non-premixed spray flamelet theory has completely ignored these effects, which can be attributed to the lack of an appropriate setup for their systematic study. 

The main objective of this work is the analysis of spray flamelet structures subject to curvature, for which a theoretical extension of the tubular counterflow flame is proposed. More specifically, the classical tubular setup is modified by including the injection of droplets from the inner cylinder and both the mathematical model presented in~\cite{Dixon-Lewis90,Dixon-Lewis91,Wang06} and the spray flamelet equations introduced in~\cite{Olguin19} are extended accordingly. The new formulations in both physical and composition space are then used to systematically analyze the ways in which curvature can modify the budgets of the flamelet equations and the stretch-induced extinction limit. The current paper contributes to advance spray flamelet theory towards a level comparable to the state of the art of non-premixed gas flamelet formulations.

%~~~~~~~~~~~~~~~~~~~~~~~~~~~~~~~~~~~~~~~~~~~~~~~~~~~~~~~~~~~~~~~
%~~~~~~~~~~~~~~~~~~~~~~~~~~~~~~~~~~~~~~~~~~~~~~~~~~~~~~~~~~~~~~~
%~~~~~~~~~~~~~~~~~~~~~~~~~~~~~~~~~~~~~~~~~~~~~~~~~~~~~~~~~~~~~~~

\section{Governing equations for tubular counterflow spray flames}
\label{sec:governingeq}
\begin{figure}[!b]
  \centering
  \includegraphics[width=0.4\linewidth]{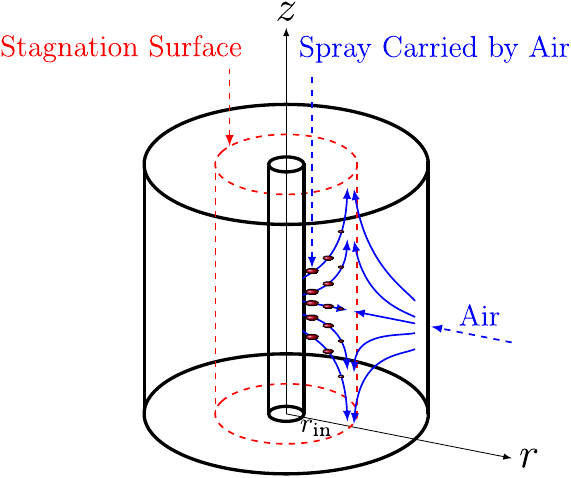}
 \caption{Schematic of a tubular counterflow spray flame ($r_{\mathrm{in}}$ denotes the radius of the inner cylinder)}.
  \label{fig:tubular}
\end{figure}

In this work, axi-symmetric steady tubular counterflow spray flames are considered, which are schematically shown in Fig.~\ref{fig:tubular}. An ethanol spray carried by air is injected from the inner cylinder and directed against a stream of air from the outer tube. 
In order to describe this setup, the gas phase equations presented in~\cite{Dixon-Lewis90,Dixon-Lewis91,Wang06} (developed for gaseous tubular flames) are extended by including the evaporation sources, which are modeled following the approach of Continillo and Sirignano~\cite{Continillo90}, and Gutheil and Sirignano~\cite{Gutheil98}. Thus, a two-dimensional Eulerian/Lagrangian formulation is obtained, which leads to a one-dimensional set of equations for the gas phase after similarity is assumed. For the liquid phase, a mono-disperse spray with spherically symmetric single-component droplets is assumed, gravitation is neglected and the droplet motion considering drag effects is formulated in a Lagrangian way~\cite{Abramzon89}. Variable transport properties and a detailed chemical reaction mechanism for ethanol/air are adopted~\cite{Gutheil98}, where the latter consists of 38 species and 337 elementary reactions~\cite{Chevalier93}. 

\subsection{Gas phase equations}

With the introduced assumptions, the transport equation for the total mass becomes
\begin{equation}
\frac{1}{r}\frac{d(r\rho u_{r})}{d r}+\rho K=\dot{S}_{v},
\label{masacil}
\end{equation}
where $u_{r}$ is the radial gas velocity, $\rho$ is the gas density, $\dot{S}_{v}$ represents the source term due to mass evaporation and $K$ is the stretch rate defined as $K=(1/A) dA/dt=d u_{z}/dz$~\cite{pitz2014}, with $A$, $u_{z}$ and $t$ denoting the flame surface, axial gas velocity and time, respectively. 

The transport equation for the momentum in the axial direction, $z$, is used to derive an equation for $K$~\cite{Dixon-Lewis90,Kalbhor22}, which yields
\begin{equation}
  \rho u_{r}\frac{dK}{dr} + \rho K^{2}=J+\frac{1}{r}\frac{d}{d r}\left(r\mu \frac{dK}{dr}\right)+\frac{\dot{S}_{mz}}{z},
  \label{momentumcil}
\end{equation}
where $\mu$ is the dynamic viscosity of the gas mixture, and $\dot{S}_{mz}$ denotes the momentum exchange with the liquid phase in $z$-direction. In this work, the axial pressure gradient, $J=-(1/z) dp/dz=\rho_{\mathrm{in}} K^{2}_{\mathrm{in}}$, (the subscript, $\mathrm{in}$, refers to the inner cylinder) is prescribed since the injection of potential flows from each cylinder is assumed~\cite{Dixon-Lewis90,Continillo90,Gutheil98,Kalbhor22}. 

The equations for the mass fractions of chemical species, $Y_k$, reads
\begin{align}
  \rho u_{r}\frac{dY_{k}}{dr}=&\frac{1}{r}\frac{d}{dr}\left(r\rho D_{k}\frac{d Y_{k}}{dr}\right)-\frac{1}{r}\frac{d}{dr}(r\tilde{V}_{kr})+\dot{\omega}_{k}+\dot{S}_{v}(\delta_{kF}-Y_{k}),
  \label{especiescil}
\end{align}
 where $D_{k}$ is the diffusion coefficient of species $k$, $\delta_{kF}$ is the Kronecker delta, where $F$ refers to the liquid fuel and $\dot{\omega}_{k}$ is the specific chemical reaction rate of species $k$. Additionally, the term $\tilde{V}_{kr}$ can be expressed as~\cite{Olguin19}
 \begin{equation}
  \tilde{V}_{kr}=-\frac{\rho Y_{k}D_{k}}{\overline{W}}\frac{d\overline{W}}{dr}-\frac{D_{Tk}}{T}\frac{dT}{dr}+\rho Y_{k}V_{kr}^{c},
  \label{vdifusion}
\end{equation}
 where $D_{Tk}$ is the thermal diffusion coefficient of species $k$, $\overline{W}$ is the mean molecular weight of the gas mixture and $V_{kr}^{c}$ is a correction term ensuring mass conservation, which reads
 \begin{equation}
  V_r^c=\sum_{k=1}^N\left(D_k \frac{Y_k}{X_k} \frac{d X_k}{d r}+\frac{D_{T k}}{\rho T} \frac{d T}{d r}\right),
\end{equation}
where $X_{k}$ is the mole fraction of species $k$.

The transport equation for the gas temperature, $T$, can be expressed as
\begin{align}
  \rho C_{p}u_{r}\frac{dT}{dr}=&\frac{1}{r}\frac{d}{dr}\left(r\lambda\frac{dT}{dr}\right)-\rho\frac{dT}{dr}\sum_{k=1}^{N}(C_{pk}V_{kr}Y_{k})+\dot{\omega}_{T}+\dot{S}_{e},
  \label{energiacil}
\end{align}
where $\lambda$ is the thermal conductivity and $\dot{\omega}_{T}$ and $\dot{S}_{e}$ are the source and sink terms due to chemical reactions and evaporation, respectively. Additionally, $C_{pk}$ and $C_{p}$ are the specific heat capacity at constant pressure of species $k$ and the gas mixture, respectively, and the diffusion velocity, $V_{kr}$, is
\begin{equation}
  V_{kr}=-\frac{D_{k}}{Y_{k}}\frac{d Y_{k}}{dr}+\frac{\tilde{V}_{kr}}{\rho Y_{k}}.\label{veldiftilde}
\end{equation}
 For the projection of the solution of the above equations into flamelet space, the mixture fraction, $Z$, must be provided. In this work, this variable is defined as a scalar satisfying the following transport equation~\cite{Olguin19,Olguin13}
\begin{equation}\rho u_{r} \frac{d Z}{d r}= \frac{1}{r} \frac{d}{d r}\left( r \rho D_{Z} \frac{d Z}{d r}\right) + \dot{S}_{v}(1-Z),\label{xitransport}\end{equation}
where $D_{Z}$ is the diffusion coefficient of the mixture fraction, assumed as $D_{Z} = \lambda /(\rho C_{p})$.

\begin{figure}[!t]
  \centering
    \includegraphics[width=0.4\linewidth]{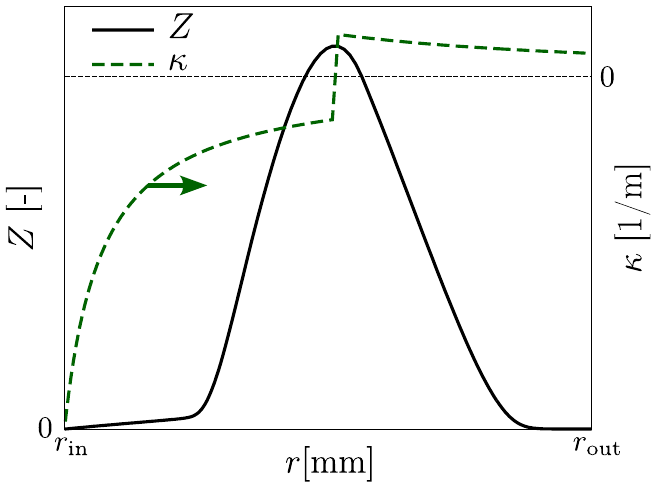}
    \caption{Schematic of $\kappa$ and $Z$ profiles in a tubular counterflow spray flame.}
    \label{fig:kappaZ}
\end{figure}

The curvature of the mixture fraction for tubular counterflow spray flames reads
\begin{equation}
  \kappa=-\nabla\cdot\left(\frac{\nabla Z}{|\nabla Z|}\right)=- \frac{1}{r}  \frac{d}{d r} \left( r \frac{\frac{dZ}{dr}}{\left|\frac{dZ}{dr}\right|} \right)=-\frac{1}{r}\mathrm{sgn}\left(\frac{dZ}{dr}\right).\label{curvature}
\end{equation}
Interestingly, according to Eq.~\eqref{curvature} and as illustrated in Fig.~\ref{fig:kappaZ}, the curvature changes its sign at the maximum of the typical non-monotonic profile of $Z$~\cite{Olguin13,Olguin142,Hollmann98,Franzelli15} due to the change of sign of $dZ/dr$. This behavior of $\kappa$ in tubular counterflow spray flames differs from their gaseous counterpart where $Z$ has a monotonic profile from $0$ to $1$ and thus, $\kappa$ keeps its sign along the domain. Additionally, it can be seen in Fig.~\ref{fig:kappaZ} that the magnitude of curvature decreases along $r$, which is readily noted in the specific profile in Eq.~\eqref{curvature}. 

The connection between the current model and classical descriptions for planar spray flames can be illustrated by expressing the diffusion terms in Eqs.~\eqref{momentumcil}, ~\eqref{especiescil}, ~\eqref{energiacil} and~\eqref{xitransport} as
\begin{align}
  \frac{1}{r}\frac{d}{dr}\left(rD_{\psi}\frac{d\psi}{dr}\right)=\frac{d}{dr}\left(D_{\psi}\frac{d\psi}{dr}\right)+\frac{D_{\psi}}{r}\frac{d\psi}{dr}\underbrace{=}_{\text{Eq.~\eqref{curvature}}}\frac{d}{dr}\left(D_{\psi}\frac{d\psi}{dr}\right)-\kappa\;\mathrm{sgn}\left(\frac{dZ}{dr}\right)D_{\psi}\frac{d\psi}{dr},
 \label{eqgen}
\end{align}
where $D_{\psi}$ is the diffusion coefficient of the general scalar $\psi$. In Eq.~\eqref{eqgen}, the second term at the right hand side comprises the effects attributable to curvature. If $r\to \infty$, thus $\kappa\to 0$ and this term can be neglected, which leads to a formulation equivalent to the one presented by Continillo and Sirignano for planar counterflow spray flames~\cite{Continillo90}. Thus, the presented set of equations is consistent with previous studies available in the literature, specially with those used in tabulation methods for spray flames~\cite{Hollmann98,Ge081,Hu172,Hu17}. 

Dirichlet boundary conditions are considered for the above equations. While the radius of the inner cylinder, $r_{\mathrm{in}}$, is imposed, the outer tube radius is assumed to be infinitely large. The gas stretch rate is imposed at both cylinders assuming a potential flow.

The gas phase equations are coupled with the liquid phase by means of the evaporation sources appearing in Eqs.~\eqref{masacil},~\eqref{especiescil},~\eqref{energiacil} and~\eqref{xitransport}. These are calculated as follows~\cite{Continillo90,Gutheil98}
\begin{equation}
    \dot{S}_{v}= \sum_{j=1}^{J} n^{d}_{j} \dot{m}^{d}_{j},\label{sv}
\end{equation}
\begin{equation}
\dot{S}_{mz}=\sum_{j=1}^{J} n_{j}^{d}\left[-m^{d}_{j} \frac{\mathrm{d} u^{d}_{z,j}}{\mathrm{d} t^{d}}+ \dot{m}^{d}_{j}(u^{d}_{z,j}-u_{z})\right]\label{sm}
\end{equation}
and
\begin{equation}
    \dot{S}_{e}=-\sum_{j=1}^{J} n_{j}^{d}\left[\dot{q}_{j}+\dot{m}^{d}_{j}\left( L_v+ \int_{T}^{T^{d}_{s,j}} C_{pF} \mathrm{d} T\right)\right],\label{se}
\end{equation}
where the subscript $j$ refers to a droplet size group, $n^d$ is the droplet number density and $\dot{m}^{d}$ is the mass vaporization rate. The droplet mass is denoted by $m^{d}$, $u_{z}^{d}$ is the axial velocity of the droplet and $t^d$ is the corresponding Lagrangian time following the droplet. The energy transferred to the droplet is denoted by $\dot{q}$ and $L_v$ is the latent heat of vaporization. The subscript $s$ (i.e. in $T^{d}_{s}$) indicates that the property is evaluated at the droplet surface.

\subsection{Liquid phase equations}
The source terms introduced in the previous section are closed with the equations for the liquid phase (droplet vaporization, motion, heating and number density) presented in~\cite{Continillo90,Gutheil98}, which are summarized here. The subscript $j$ is omitted for clarity. 

The equation for the mass vaporization rate, $\dot{m}^{d}$, in Eqs.~\eqref{sv},~\eqref{sm}, and~\eqref{se} reads~\cite{Continillo90}
\begin{equation}
\dot{m}^{d}=2 \pi R \rho_{f} D_{f} \widetilde{\mathrm{Sh}} \ln \left(1+B_{M}\right),\label{mdot}
\end{equation}
where $R$ is the droplet radius. The subscript $f$ refers to the properties in the film around the droplet (computed using the 1/3 rule~\cite{Hubbard75}). The modified Sherwood number, $\widetilde{\mathrm{Sh}}$, and the Spalding mass transfer number for the droplet, $B_M$, are~\cite{Continillo90,Abramzon89}
\begin{equation}
\widetilde{\mathrm{Sh}}=2+\frac{([1+\mathrm{Re}\mathrm{Sc}]^{1/3}[\max(1,\mathrm{Re})]^{0.077}-1)}{(1+\mathrm{B}_M)^{0.7}\log(1+\mathrm{B}_M)}\mathrm{B}_M
\end{equation}
and
\begin{equation}
\mathrm{B}_M=\frac{Y_{Fs}-Y_F}{1-Y_{Fs}},
\end{equation}
respectively, where $\mathrm{Re}$ and $\mathrm{Sc}$ denote the Reynolds and Schmidt numbers, respectively. Additionally, the expression for the droplet mass, $m^{d}={4}\pi R^3\rho^{d}/3$, can be derived in time and inserted into Eq.~\eqref{mdot} to obtain an equation for the droplet radius, which yields
\begin{equation}
\frac{\mathrm{d} }{\mathrm{d} t^d}\left(\frac{4}{3}\pi R^3\rho^{d}\right)=-2 \pi R \rho_{f} D_{f} \widetilde{\mathrm{Sh}} \ln \left(1+B_{M}\right).
\end{equation}

In order to obtain the droplet position, $\textbf{x}^{d}$, and velocity, $\textbf{u}^{d}=\mathrm{d} \textbf{x}^{d}/\mathrm{d} t^d$, the equation for the droplet motion is used, which reads~\cite{Continillo90,Gutheil98}
\begin{equation}
 m^{d}\frac{\mathrm{d} \mathbf{u}^{d}}{\mathrm{d} t^d}=\pi R^2 \frac{1}{2} \rho\left(\mathbf{u}-\mathbf{u}^d\right)\left|\mathbf{u}-\mathbf{u}^{d}\right| C_D,
 \end{equation}
 where $C_{D}=12\nu/(|\mathbf{u}-\mathbf{u}^{d}|R)$ is the drag coefficient, calculated assuming Stokes flow and $\nu$ is the kinematic viscosity of the gas.

For the closure of Eq.~\eqref{se}, $\dot{q}$ is computed as~\cite{Continillo90,Gutheil98} 
 \begin{equation}
\dot{q}=\dot{m}^{d}\left(\frac{C_{pf}(T-T^{d}_{s})}{B_T}-L_{v}\right),
\end{equation}
with $B_{T}$ denoting the Spalding heat transfer number, which yields~\cite{Continillo90,Abramzon89}
\begin{equation}
\mathrm{B}_T=(1+\mathrm{B}_M)^\phi-1,
\end{equation}
where
\begin{equation}
\phi=\frac{C_{pl}\widetilde{\mathrm{Sh}}}{C_{pf}\widetilde{\mathrm{Nu}}}\frac{1}{\mathrm{Le}}.
\end{equation}
Here, the modified Nusselt number is calculated with
\begin{equation}
\widetilde{\mathrm{Nu}}=2+\frac{([1+\mathrm{Re}\mathrm{Pr}]^{1/3}[\max(1,\mathrm{Re})]^{0.077}-1)}{(1+\mathrm{B}_T)^{0.7}\log(1+\mathrm{B}_T)}\mathrm{B}_T,
\end{equation}
where $\mathrm{Pr}$ denotes the Prandtl number.

In order to obtain the temperature of the liquid, $T^{d}$, the droplet heating equation is used, which reads
 \begin{equation}
 \frac{\partial T^{d}}{\partial t^{d}}= \frac{1}{\zeta^2} \frac{\partial}{\partial \zeta}\left(\alpha^{d}\zeta^2 \frac{\partial T^{d}}{\partial \zeta}\right),\label{dropheating}
\end{equation}
where $\alpha^{d}$ is the thermal diffusivity of the liquid, respectively, and $\zeta$ is the radial coordinate inside the droplet, $0\leq \zeta\leq R$. The boundary conditions for Eq.~\eqref{dropheating} are 
\begin{equation}
\left.\frac{\partial T^{d}}{\partial \zeta}\right|_{\zeta=0}=0
\end{equation}
and 
\begin{equation}
\left.\frac{\partial T^{d}}{\partial \zeta}\right|_{\zeta=R}=\frac{\dot{q}}{4 \pi R^2 \alpha^{d} \rho^{d} C_{p F}}.
\end{equation}
Finally, the particle conservation equation for the droplet number density, $n^d$, reads~\cite{Continillo90,Gutheil98}
\begin{equation}
\nabla\cdot(n^d\textbf{u}^{d}) = \dot{S}_{n},
\label{particl}
\end{equation}
where $\dot{S}_{n}$ is a source term to describe the change in $n^d$ if a droplet reverses or oscillates~\cite{Gutheil98}. An analytical solution can be obtained for $n^{d}$ in the present configuration by applying the procedure of Continillo and Sirignano~\cite{Continillo90} in cylindrical coordinates, which yields

\begin{equation}
n^{d}=\frac{(n^{d}u^{d}_{r}z^{d}r^{d})_{t=0}}{u^{d}_{r}z^{d}r^{d}},
\label{densidadgotas}
\end{equation}
where $r^{d}$ and $z^{d}$ denote the radial and axial droplet position, respectively, and $u_{r}^{d}$ is the radial component of droplet velocity. It is noteworthy that the difference between this solution and the presented one for planar flame structures~\cite{Continillo90,Gutheil98} relies on $r^{d}_{t=0}/r^{d}$. This factor can be expressed in terms of the curvatures as
\begin{equation}
  \frac{r^{d}_{t=0}}{r^{d}}\underbrace{=}_{\text{Eq.~\eqref{curvature}}}\frac{-\kappa}{-\kappa_{\mathrm{in}}}\frac{\mathrm{sgn}\left(dZ/dr\right)_{\mathrm{in}}}{\mathrm{sgn}\left(dZ/dr\right)}=\frac{|\kappa|}{|\kappa_\mathrm{in}|},\label{r1_k}
\end{equation}
where the absolute values of $\kappa$ appear because $dZ/dr$ and $\kappa$ always have opposite signs (cf. Fig.~\ref{fig:kappaZ}) and are multiplied by a minus sign. In Eq.~\eqref{r1_k}, it can be noted that the decreasing profile of $|\kappa|$ (cf. Fig.~\ref{fig:kappaZ}) contributes to the decrease of $n$ in Eq.~\eqref{densidadgotas} and thus in the evaporation sources.

With the equations for both gas and liquid phases provided, the employed numerical procedure is the same as in~\cite{Continillo90,Gutheil98}. The results obtained with the presented mathematical model are projected into composition space for the study of the flamelet structures. This includes the evaluation of the different terms in the flamelet equations, which are shown in the next section.
%%%%%%%%%%%%%%%%%%%%%%%%%%%%%%%%%%%%%%%%%%%%%%%%%%%%%%%%%%%%%%%%%%%%%%%%%%%%%%%%%%%%%%%%%%%%%%%%%%%%%%%%%%%%%%%%%%%%%%%%%%%%%%%%%%%%%%%%%%
\section{Spray flamelet equations considering curvature}\label{flameleteq}

We start introducing the following flamelet transformation
\begin{equation}
 \frac{d(\cdot)}{dr}= \frac{dZ}{dr}\frac{d(\cdot)}{dZ}
 \label{drdxi},
\end{equation}
which, after inserted into Eqs.~\eqref{especiescil} and~\eqref{energiacil}, leads to the flamelet equations for temperature and chemical species. These can be expressed as
\begin{equation}
\underbrace{\rho D_{Z} g_{Z}^{2}\frac{d^{2}T}{d Z^{2}}}_{\text{Diffusion}}\!+\!\underbrace{\frac{\dot{\omega}_{T}}{C_{p}}}_{\substack{\text{Ch.} \\ \text{Source}}}\!\underbrace{-\;\rho f_{T}\frac{d T}{d Z}}_{\text{Convection}}\!+\!\underbrace{\frac{\dot{S}_{e}}{C_{p}}}_{\substack{\text{Evap.} \\ \text{Sink}}}\!+\!\underbrace{\Omega_{T}}_{\substack{\text{Other} \\ \text{Terms}}}=0\label{flameletT}
\end{equation}
and
\begin{equation}
\underbrace{\frac{\rho D_{Z}g^{2}_{Z}}{\mathrm{Le}_{k}}\frac{d^{2} Y_{k}}{d Z^{2}}}_{\text{Diffusion}}+\underbrace{\dot{\omega}_{k}}_{\substack{\text{Ch.} \\ \text{Source}}}\underbrace{-\;\rho f_{k}\frac{d Y_{k}}{dZ}}_{\text{Convection}}+\underbrace{\Omega_{Y_{k}}}_{\substack{\text{Other} \\ \text{Terms}}}=0,\label{flameletY}
\end{equation}
respectively, where $g_{Z}=|dZ/dr|$ has been introduced and $\mathrm{Le}_{k}=\lambda/(\rho C_{p}D_{k})$ corresponds to the Lewis number of chemical species $k$. 

In Eqs.~\eqref{flameletT} and~\eqref{flameletY}, the first and second terms denote the classical effects of diffusion and chemical reactions. On the other hand, the third terms formally represent convection effects in composition space, where $f_{T}$ and $f_{k}$ are
\begin{equation}
f_{T}=\underbrace{\frac{\dot{S}_{v}}{\rho}(1-Z)}_{\text{Evaporation}}\label{CT}
\end{equation}
and
\begin{align}
f_{k}=f_{T}+\underbrace{\frac{g_{Z}}{\rho}\frac{d}{d Z}\left(\left(
  1-\frac{1}{\mathrm{Le}_{k}}\right)\rho D_{ Z} g_{Z}\right)}_{\text{Differential Diffusion}}\underbrace{-\kappa\left(1-\frac{1}{\mathrm{Le}_{k}}\right) D_{ Z}g_{Z}}_{\text{Curvature}},\label{componentsvel}
\end{align} 
respectively. It can be noted that $f_{T}$ only contains evaporation effects~\cite{Olguin13}, which are also accounted in the first term in Eq.~\eqref{componentsvel}. Thus, evaporation modifies the factors $f_{T}$ and $f_{k}$ in both flamelet equations in the same way. The second term in $f_k$ accounts for differential diffusion~\cite{Pitsch98}, while the third one contains the curvature effects~\cite{Scholtissek16}. 

Finally $\Omega_{T}$ and $\Omega_{k}$ in Eqs.~\eqref{flameletT} and~\eqref{flameletY} contain terms that are expected to be small. They read
\begin{equation}
  \Omega_{T}=\frac{\rho D_{Z}g^{2}_{Z}}{C_{p}} \frac{d C_{p}}{d Z}\frac{d T}{d Z}-\frac{\rho g_{Z}}{C_{p}}\sum_{k=1}^{N}C_{p,k}Y_{k}V_{kZ}\frac{d T}{d Z}
\end{equation}
and
\begin{equation} \Omega_{Y_{k}}=\dot{S}_{v}(\delta_{kF}-Y_{k})+\kappa\tilde{V}_{k Z}g_{Z}-\frac{d \tilde{V}_{k Z}g_{Z}}{d Z}g_{Z},
\end{equation}
respectively, where the diffusion velocity in composition space, $V_{kZ}$, is ~\cite{Olguin19} 
\begin{equation}
V_{kZ}=-\frac{D_{Z}}{Y_{k}\mathrm{Le}_{k}}\frac{dY_{k}}{dZ}+\frac{\tilde{V}_{kZ}}{\rho Y_{k}}, 
\end{equation}
where
\begin{align}  \tilde{V}_{kZ}=-\frac{\rho Y_{k}D_{k}}{\overline{W}}\frac{d\overline{W}}{dZ}-\frac{D_{Tk}}{T}\frac{dT}{dZ}+\rho Y_{k}\sum_{k=1}^N\left(D_k \frac{Y_k}{X_k} \frac{d X_k}{d Z}+\frac{D_{T k}}{\rho T} \frac{d T}{d Z}\right).
\end{align}

The different contributions in Eqs.~\eqref{flameletT},~\eqref{flameletY} and~\eqref{componentsvel} are evaluated in the results' section to study how they are modified by curvature.

\section{Results}

The introduced mathematical framework will be employed now to study the effects of curvature on spray flamelet structures. The analysis begins with the consideration of three selected counterflow flames subject to different levels of curvature, which are examined in both physical and mixture fraction space. The main findings of this study are then generalized by means of a comprehensive parametric analysis in which the effects of increasing stretch and curvature on flamelet extinction are emphasized.
%%%%%%%%%%%%%%%%%%%%%%%%%%%%%%%%%%%%%%%%%%%%%%%%%%%%%%%%%%%%%%%%%%%%%%%%%%%%%%%%%%%%%%%%%%%%%%%%%%
\subsection{Curvature effects on spray flamelet structures}\label{abovezst} 
\begin{figure*}[!b]
 \centering  \includegraphics[width=1\linewidth]{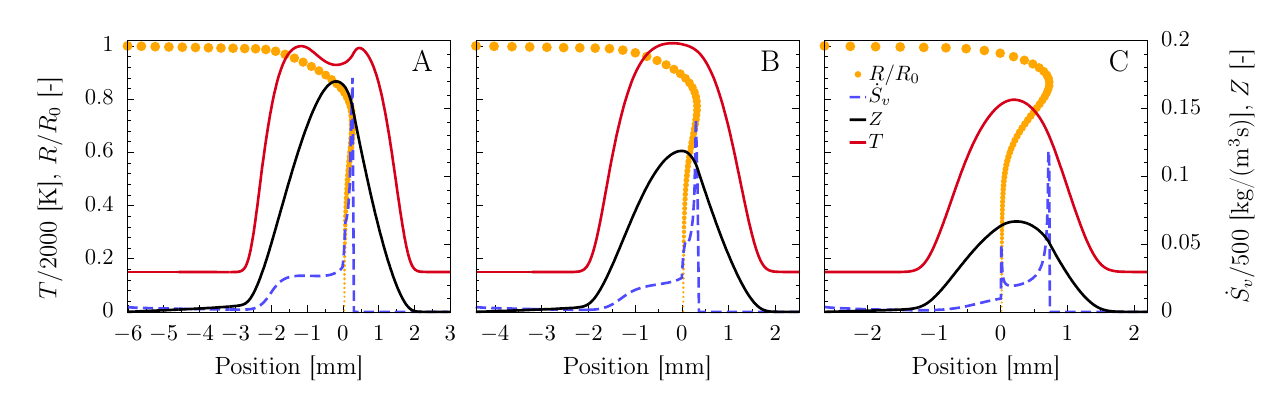}
   \caption{Normalized droplet radius, evaporation mass source, mixture fraction and gas temperature for the selected flames ($0$ in the abscissa represents the stagnation surface).}
\label{fig:Structuresphys}
\end{figure*}
 \begin{figure}[!hb]
     \centering

\includegraphics[width=0.46\linewidth]{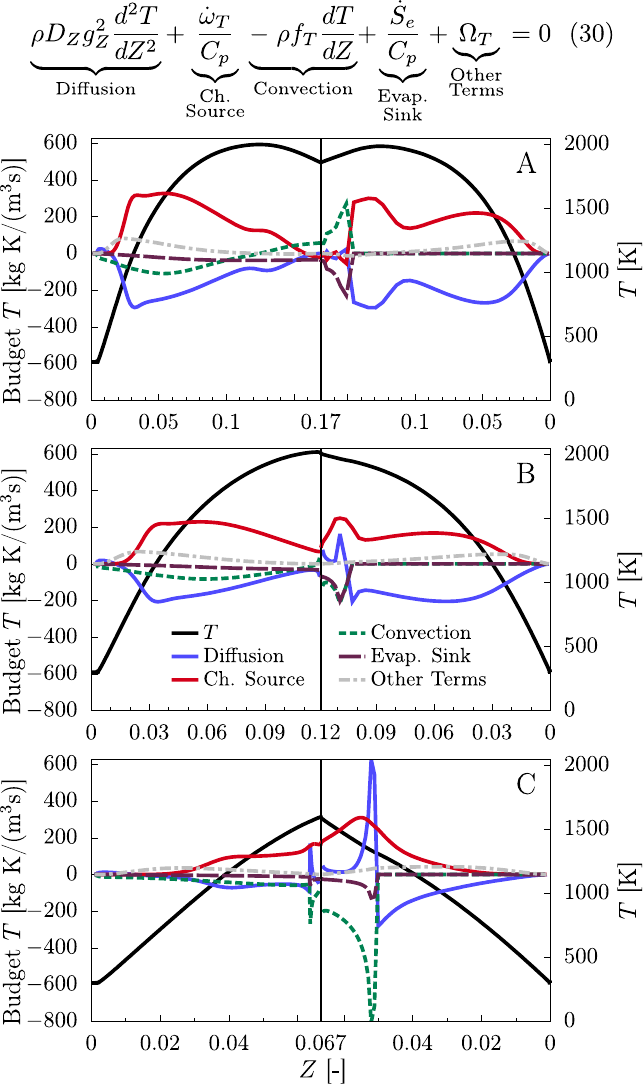}
\includegraphics[width=0.444\linewidth]{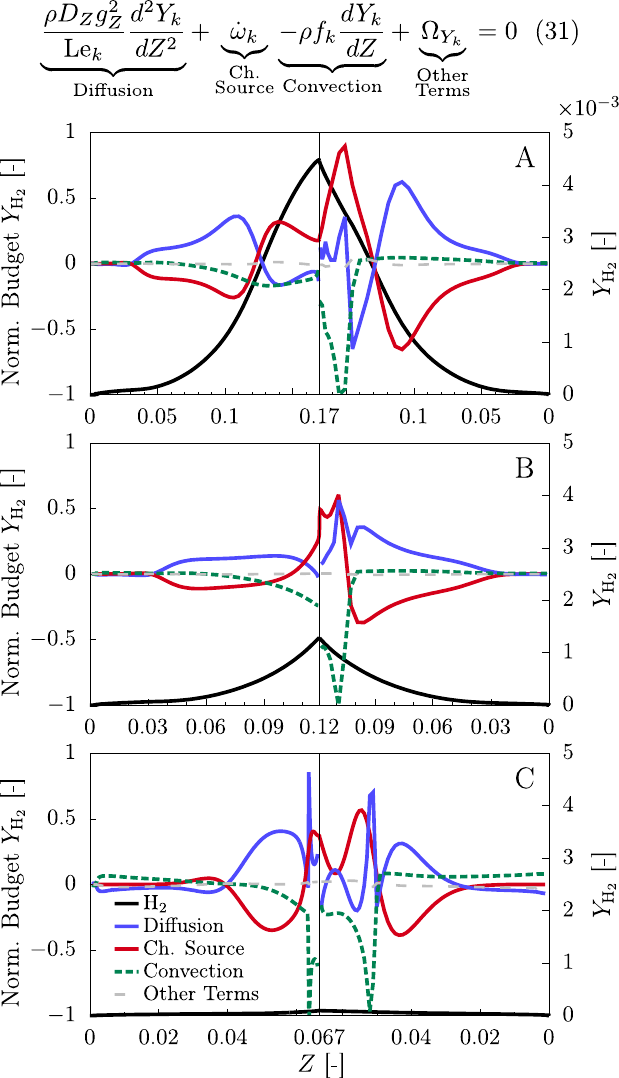}
     \caption{Budgets of the flamelet equations for $T$ (left) and $Y_{\mathrm{H}_{2}}$ (right) for the selected flames. For the latter, all budgets are normalized by their largest term. The vertical line denotes $Z_\mathrm{max}$.}
     \label{fig:BudgetsT}
 \end{figure}
We start considering three particular flames, A, B and C, with different curvatures at the inner cylinder, $\kappa_{\mathrm{in}}=0$/m, $-200$/m and $-925$/m. These particular values are selected because they represent flames with zero, moderate and the maximum possible magnitude of $\kappa$, respectively, before extinction is reached. As already explained, these flames are established by injecting a spray carried by air from the inner cylinder, which is directed against an air stream coming from the outer tube (cf. Fig.~\ref{fig:tubular}). For all cases, the gas temperature at both sides is $300$~K, the gas velocity at the spray side is $1.2$~m/s, and $K_{\mathrm{in}}=300$/s. The initial droplet velocity and temperature match the gas phase conditions at the boundary, while the initial droplet radius is $25$~\textmu m and the liquid fuel/air equivalence ratio, $E$, is unity.

Figure~\ref{fig:Structuresphys} displays the profiles of the normalized droplet radius, $R/R_0$, gas temperature, evaporation mass source and mixture fraction, for the three flames under consideration. It is found that the peak value of $\dot{S}_{v}$ is considerably reduced by increasing curvature, which significantly affects the flamelet structures. In particular, the three selected curvatures lead to very different mixture fraction and temperature profiles: For Flame A, a mixture fraction peak well above its stoichiometric value, $Z_{\mathrm{st}}\approx0.11$, is observed, which leads to two different reaction zones (related to corresponding peaks of the temperature profile) one at each side of the configuration. For Flame B, on the other hand, these two reaction zones are merged and stoichiometry is reached only at the point of $Z_\mathrm{max}$. Consequently, the temperature profile has a single peak at this location. Finally, Flame C corresponds to a lean flame, which presents a maximum value of the mixture fraction below stoichiometry and a temperature peak considerably lower than for Flames A and B. 

\begin{figure}[!b]
    \centering
\centering
\includegraphics[width=1\linewidth]{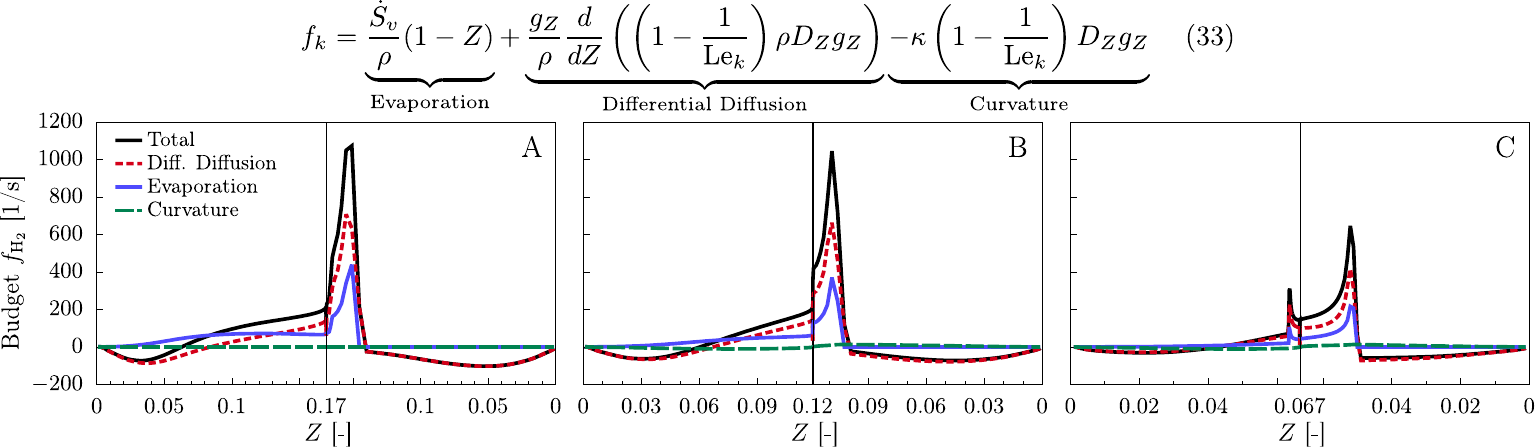}
  \caption{Budgets of $f_{\mathrm{H}_{2}}$ for the selected flames. The vertical line denotes $Z_\mathrm{max}$.}
 \label{fig:budgetsCh2}
\end{figure}
The consequences of the modifications of the temperature profile can be better understood in terms of the budgets of the $T$-spray flamelet equation, which are obtained by projecting the physical space results into mixture fraction space (see left of Fig.~\ref{fig:BudgetsT}). For all flames, a reaction-diffusion dominated structure is observed in almost the entire domain, with relevant evaporation-associated effects only at the point of maximum $\dot{S}_v$. At this location, the most notable differences between Flames A, B and C are the convective and diffusive terms. In particular, the convective term exhibits a positive value for Flame A, and its sign changes for Flames B and C. Since $f_{T}=\dot{S}_{v} (1-Z)/\rho$ is positive along the entire domain, this change of sign is attributed to the corresponding change of $dT/dZ$ (see the $T$-profile on the left of Fig.~\ref{fig:BudgetsT}). In response to this, the diffusion term becomes positive and, together with the chemical source term, balances convection for Flame C.
As pointed out before, $\kappa_\mathrm{in} = -925$ /m is the highest value for which a steady flamelet structure could be obtained and any further increase of this quantity would lead to flame extinction, which can be attributed to the fact that diffusion would not be able to compensate anymore the sink of energy generated by the composition-space convective term. It is interesting to note that this mechanism, which will be analyzed in more detail in the next section, considerably differs from the classical extinction phenomenon typically observed in gas flames. In general, gas flamelet extinction occurs due to chemical reactions not being able to compensate the diffusion associated with high values of the scalar dissipation rate~\cite{Peters84}. From this analysis, which has been conducted for temperature ($\mathrm{Le}=1$), it is concluded that curvature strongly impacts spray flamelet structures.

For the next analysis, $\mathrm{H}_2$ is selected as a reactive scalar specially sensitive to curvature ($\mathrm{Le}_{\mathrm{H}_2}\approx 0.3\ll 1$). The normalized budgets of the $Y_{\mathrm{H}_2}$-flamelet equation are shown on the right side of Fig.~\ref{fig:BudgetsT}, which immediately reinforce the relevance of the convective term at the location of the maximum value of $\dot{S}_v$. As readily noted, at that point, this term is balanced by both diffusion and chemical reactions for all flames under consideration. This occurs because the value of $dY_{\mathrm{H}_2}/dZ$ is always positive there. Despite the qualitative similarities between the budgets of Flames A, B and C, important quantitative differences are found, which are directly attributable to evaporation and differential diffusion. Further, the budgets of $f_{\mathrm{H}_{2}}$ show that the term $\kappa(1-1/\mathrm{Le_{k}})D_{Z}g_{Z}$ is negligible (see Fig.~\ref{fig:budgetsCh2}). This minor importance can be attributed to both the low order of magnitude of $D_{Z}$ ($\sim10^{-4}$, not shown) and the limited range of $\kappa$ available with the tubular configuration~\cite{Xuan14,Scholtissek16,Bottler21}. The latter can be explained by the already discussed monotonic decrease of $\kappa$ in the proposed setup (cf. Fig~\ref{fig:kappaZ}). Based on these results, it is concluded that the spray flamelet structures studied in this work are considerably more sensitive to curvature effects than similar hydrocarbon flames analyzed in the literature~\cite{Scholtissek16}. However, even for highly diffusive species such as H$_2$ ($\mathrm{Le}_{\mathrm{H}_2}\ll1$), curvature effects are not directly introduced into the corresponding flamelet equation by terms explicitly accounting for this quantity, but rather by evaporation and differential diffusion.
%%%%%%%%%%%%%%%%%%%%%%%%%%%%%%%%%%%%%%%%%%%%%%%%%%%%%%%%%%%%%%%%%%%%%%%%%
\subsection{Analysis of the extinction limit}\label{sec:global}
In this subsection the effects of curvature on the stretch-induced extinction of spray flamelets are studied. For this, a parametric analysis is performed, where several counterflow flames with curvatures ranging from $0$/m to $-1000$/m are carried to extinction by systematically increasing the stretch rate, $K_\mathrm{in}$, starting from 55/s. All other boundary conditions correspond to the ones established in Section~\ref{abovezst}.
   \begin{figure}[!b]
   \centering
\includegraphics[width=0.47\linewidth]{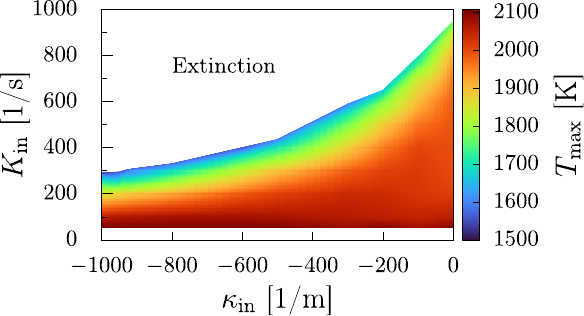}
\includegraphics[width=0.47\linewidth]{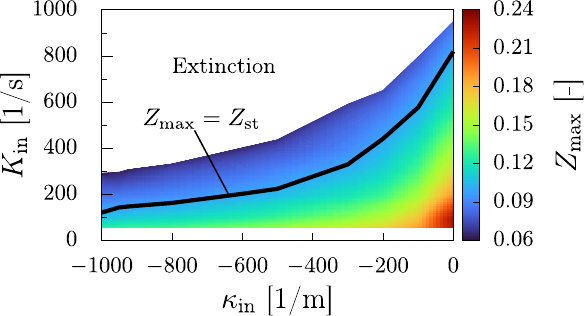}
\caption{$T_{\mathrm{max}}$ (left) and $Z_{\mathrm{max}}$ (right) for different curvature and stretch rate conditions. The white region represents extinguished flames. The black line indicates the combinations of stretch and curvature for which $Z_{\mathrm{max}}=Z_{\mathrm{st}}\approx0.11$.}
   \label{fig:Tmax2D} 
 \end{figure}
\begin{figure}[!t]
\centering
\includegraphics[width=0.5\linewidth]{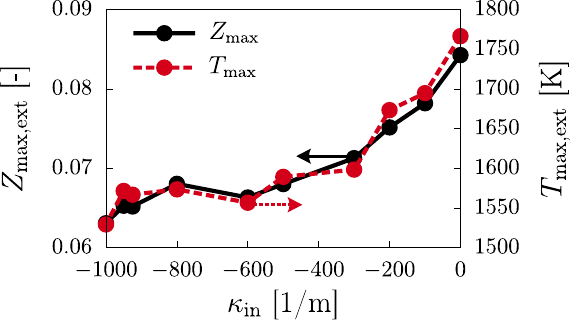}
 \caption{Maximum mixture fractions and temperatures at stretch-induced extinction conditions for different values of curvature.}
\label{fig:TmaxZmaxext}
 \end{figure}
\begin{figure}[!t]
\centering
\includegraphics[width=1\linewidth]{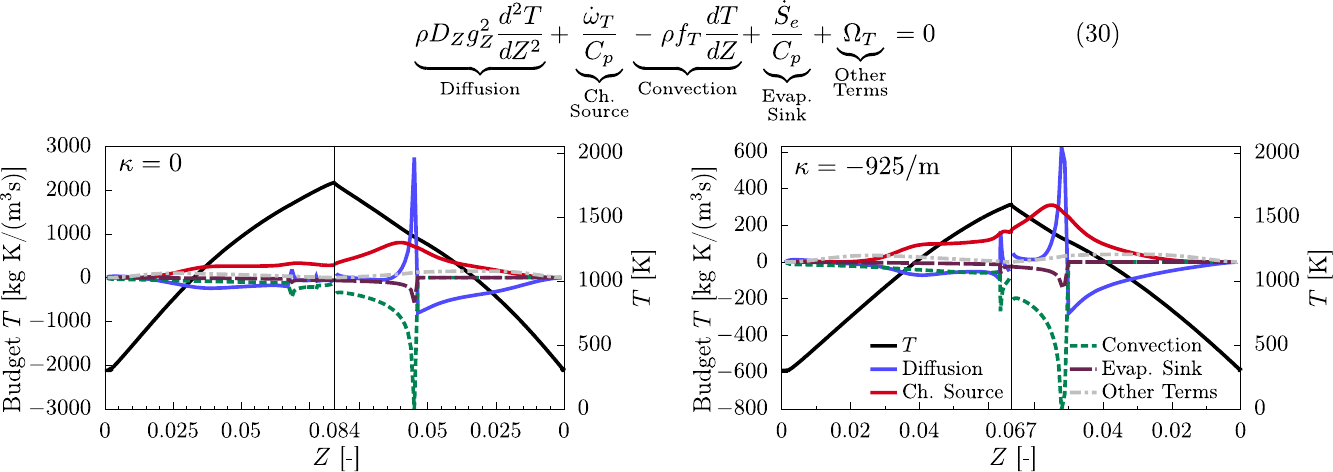}
  \caption{Budgets of $T$-flamelet equations for flames close to stretch-induced conditions. The vertical line denotes $Z_\mathrm{max}$.}
  \label{fig:budgets_ext}
\end{figure}

Figure~\ref{fig:Tmax2D} shows the maximum gas temperature and mixture fraction ($T_{\mathrm{max}}$ and $Z_{\mathrm{max}}$, respectively), as a function of stretch and curvature. It is observed that the extinction stretch value, $K_{\mathrm{ext}}$, significantly decreases with curvature. Additionally, two zones can be clearly identified on the left of Fig.~\ref{fig:Tmax2D}. The first one corresponds to a high temperature region, which is located towards low values of stretch. On the other hand, a low $T_{\mathrm{max}}$ zone can be distinguished very close to the extinction border. In the corresponding distribution of $Z_{\mathrm{max}}$ (see right of Fig.~\ref{fig:Tmax2D}), it can be seen that the two identified regions are separated by the $Z_{\mathrm{max}}=Z_{\mathrm{st}}$-isoline (black). Above this line, $Z_{\mathrm{max}}<Z_{\mathrm{st}}$ and the temperature drastically decreases in the same way it has been illustrated by Flame C in the previous section. Of course, this phenomenon has not been observed in gas tubular counterflow flames~\cite{pitz2014,Xuan14,Scholtissek16} since pure fuel is injected from one side and, in consequence, $Z_{\mathrm{max}}=1$, which ensures the achievement of stoichiometric conditions.
 
Figure~\ref{fig:TmaxZmaxext} shows the values of $T_{\mathrm{max}}$ and $Z_{\mathrm{max}}$ at $K_{\mathrm{ext}}$ as a function of $\kappa_{\mathrm{in}}$. It can be seen that flames with higher curvatures extinguish at lower temperatures and farther from stoichiometry than planar flames. In other words, planar flamelets can be rapidly extinguished once $Z_{\mathrm{max}}$ decreases below its stoichiometric value, even when the temperature is still relatively high. Curved flamelets, on the other hand, can be established for values of $Z_{\mathrm{max}}$ way below $Z_{\mathrm{st}}$, exhibiting maximum temperatures up to about $240$ K lower than their planar counterpart.

Finally, the budgets of the extended temperature spray flamelet equation close to stretch-induced extinction are displayed in Fig.~\ref{fig:budgets_ext} for two different curvatures. It is clearly seen that, while quantitative differences exist, the extinction mechanism in both cases is the same: Evaporation-induced convection in mixture fraction space becomes a major energy sink, which can no longer be compensated by diffusion and the chemical reaction source term. This mechanism is the same observed in the previous section for Flame C, and, as pointed out before, it completely differs from the one typically observed in gas flamelet theory, where extinction occurs due to chemical reactions not being able to balance diffusion. 
%~~~~~~~~~~~~~~~~~~~~~~~~~~~~~~~~~~~~~~~~~~~~~~~~~~~~~~~~~~~~~~~
\section{Conclusions}\label{sec:con}
In this work, spray flamelet structures subject to curvature have been systematically analyzed, emphasizing how this quantity modifies the different terms appearing in the flamelet equations and the stretch-induced extinction limit. For this, a theoretical extension of the tubular counterflow configuration has been proposed to allow spray injection from the inner tube, and both the formulation presented in~\cite{Dixon-Lewis90,Dixon-Lewis91,Wang06} and the flamelet equations introduced in~\cite{Olguin19} have been extended accordingly. Making use of these formulations, several ethanol/air tubular counterflow flames were simulated and analyzed in both physical and composition space. The results revealed that increasing curvature leads to important modifications of the resulting flamelet structures, which can be attributed to the
effect of curvature on the evaporation profiles. Further, a parametric study was performed, showing that higher curvature leads to significantly lower stretch-induced extinction limits. This was related to a reduction of the maximum mixture fraction to values farther below its stoichiometric value.

Finally, the analysis of the budgets of the spray flamelet equations including curvature effects revealed that the mechanism by which flame extinction is reached in the tubular counterflow configuration significantly differs from the one observed in gas flames. In particular, it has been found for spray flamelets that evaporation-induced convection in mixture fraction space acts as the dominant energy sink leading to flame quenching, which can no longer be compensated by chemical reaction and diffusion. In contrast, for gaseous flamelets extinction occurs when chemical reactions can no longer balance diffusion.
%~~~~~~~~~~~~~~~~~~~~~~~~~~~~~~~~~~~~~~~~~~~~~~~~~~~~~~~~~~~~~~~
\section*{Acknowledgments}

The authors acknowledges funding from Proyecto Interno USM PI\_LIR\_2022\_15, FH and FR thank ANID (Chile) for financial support through the Doctorate Scholarships 21201826 and 21212269, respectively. FR is also grateful for the financial contribution from the PIIC-USM grant. AS and CH acknowledge funding by the Deutsche Forschungsgemeinschaft (DFG, German Research Foundation) - Projektnummer 325144795.

\bibliography{references}
\end{document}